# Automated Water Irrigation System

**David Veksler and Matthew Okner**
West Valley College, Saratoga, CA 95070 USA

**ABSTRACT** This paper presents the design and implementation of an automated water irrigation system aimed at optimizing plant care through precision moisture monitoring and controlled water delivery. The system uses a capacitive soil moisture sensor, an ADC (analog-to-digital converter), and a relay-driven water pump to ensure plants receive adequate hydration based on real-time data. In addition, this work aims to build on existing applications for Raspberry Pi (4B) and Arduino-based automatic irrigation systems by integrating advanced calibration methods, employing optimized algorithms, and introducing new technologies to further enhance overall system efficiency and reliability.

## GITHUB PROJECT REPOSITORY

https://github.com/maokner/irrigationproject

## I. INTRODUCTION

Water conservation is a growing global concern, particularly in agriculture, which accounts for a significant portion of freshwater consumption. An automated irrigation system offers a sustainable solution by delivering precise amounts of water only when necessary. This project aims to design a compact, low-cost, and user-friendly irrigation system that can monitor soil moisture levels and water plants autonomously. The proposed system leverages modern hardware and software technologies to achieve efficiency and reliability, making it suitable for various environments, from small gardens to large-scale farms.

## II. DESIGN OVERVIEW

The automated irrigation system is designed around three primary components: sensing, control, and actuation, all working together to ensure efficient water management. The sensing component relies on a capacitive soil moisture sensor placed directly in the soil. Unlike resistive sensors that are prone to corrosion, the capacitive sensor uses changes in capacitance to measure the soil's moisture content. This provides a more durable and accurate way to assess the hydration level of the soil, generating an analog signal proportional to the moisture level.

The control system is built around the ADS 1115, an analog-to-digital sigma-delta converter (a Texas Instrument ADC) that transforms the analog signals from the moisture sensor into digital values readable by the Raspberry Pi. These values are processed by the Raspberry Pi to determine whether the soil moisture has dropped below a predefined threshold. If this threshold is breached, the system activates a relay module. The relay serves as a switch, enabling or disabling the water pump based on the Raspberry Pi's instructions. This modular design ensures that the system is both responsive and adaptable to varying environmental conditions.

Finally, the actuation system consists of a 5V relay module connected to a water pump powered by a battery. When the relay is triggered, it completes the circuit to power the pump, allowing water to be delivered to the soil. The pump dispenses water for a duration calculated by the system's software, ensuring that only the necessary amount of water is provided. This approach minimizes water wastage and prevents overwatering, maintaining an optimal moisture level for plant health. Together, the sensing, control, and actuation components form an integrated system capable of efficiently managing irrigation with minimal user intervention.

## III. ALGORITHM

The system algorithm follows a structured process to ensure efficient and precise irrigation. It begins with the moisture reading phase, where the soil moisture sensor collects real-time data about the soil's hydration level. This data is sent as an analog signal to the ADS1115, which converts it into a digital value that can be processed by the Raspberry Pi. The moisture level is then compared against a predefined threshold value, which represents the minimum acceptable soil moisture content. This threshold is chosen to prevent plant dehydration while avoiding overwatering.

If the measured moisture level falls below the threshold, the system initiates the watering process. The Raspberry Pi sends a signal to the relay module, which activates the water pump. The amount of water dispensed is calculated based on the severity of the moisture deficit and the user-defined settings. This ensures that only the necessary quantity of water is delivered to the soil, optimizing water usage. During this phase, the system also updates its internal table to reflect that watering is in progress, providing real-time status updates to the user interface.

After the watering process is complete, the system enters a feedback loop. The soil moisture level is rechecked to confirm that it has been restored to the desired threshold. If the moisture level is still insufficient, the system repeats the watering process until the target level is reached. Additionally, the system automatically schedules the next moisture check to occur after a predefined interval, ensuring continuous monitoring.

## IV. CAPACITIVE MOISTURE SENSOR

The capacitive soil moisture sensor works by measuring changes in capacitance caused by varying moisture levels in the soil. The capacitance of a capacitor depends on the material between its plates, which is given by the formula[1]:

$$C = \frac{\varepsilon_r \varepsilon_0 A}{d}$$

In this equation, $\varepsilon_r$ is the material's dielectric constant (soil), $\varepsilon_0$ is the permittivity of free space, A is the plate area, and d is the distance between the plates. When the soil contains more water, the dielectric constant increases because water has a much higher value ($\varepsilon_r \approx 80$) than dry soil ($\varepsilon_r \approx 2-4$)[2]. This increase in the dielectric constant raises the capacitance, allowing the sensor to detect changes in moisture.

To make the sensor accurate, we calibrated it using two reference points: dry soil and saturated soil. For the dry soil calibration, the sensor was placed in completely dry soil, and its output was recorded as the 0% moisture level. For the saturated soil calibration, the sensor was placed in fully soaked soil, and its output was recorded as the 100% moisture level. These two points allowed us to map the sensor's readings to a percentage scale for soil moisture.

## V. SOFTWARE

The software ties the entire irrigation system together, handling data collection, decision-making, and user interaction. A Python backend, built with Flask, runs on the Raspberry Pi to process moisture data from the sensor and decide when to activate the water pump. The system includes a web interface created with HTML, CSS, and JavaScript, which allows users to view moisture levels, system status, and control watering manually. The interface updates automatically with real-time data, ensuring users always have accurate information. This software design ensures the system is easy to use while managing the irrigation process efficiently.

The system operates on a dynamic push-pull sequence. When a user initiates an action, such as checking the soil moisture or watering the plant, the website sends the command to the backend, which processes it and interacts with the sensor or relay as needed. Updated data, such as the new moisture level or watering status, is then retrieved from the backend and displayed on the website.

The backend is designed to work in tandem with a monitoring thread that runs continuously in the background. This thread checks the soil moisture level at regular intervals (currently set to 30 minutes as per Kawahara's Study[3]) and determines whether irrigation is required. If the moisture level falls below the predefined threshold, the system automatically waters the plant and updates the website with the relevant details.

## VI. Website Diagram

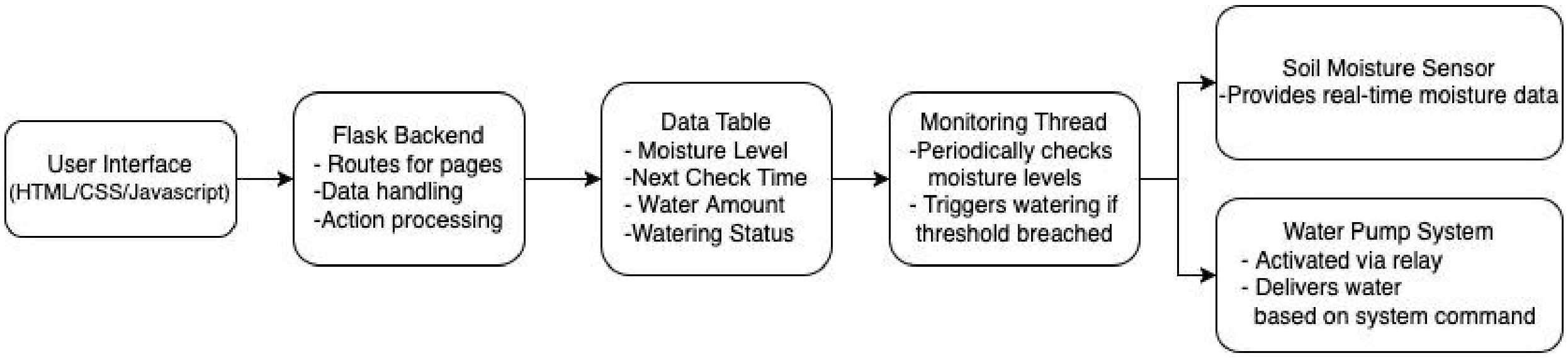

## GITHUB PROJECT REPOSITORY

https://github.com/maokner/irrigationproject

## VII. IMPLEMENTATION, DEMONSTRATION AND DESIGN

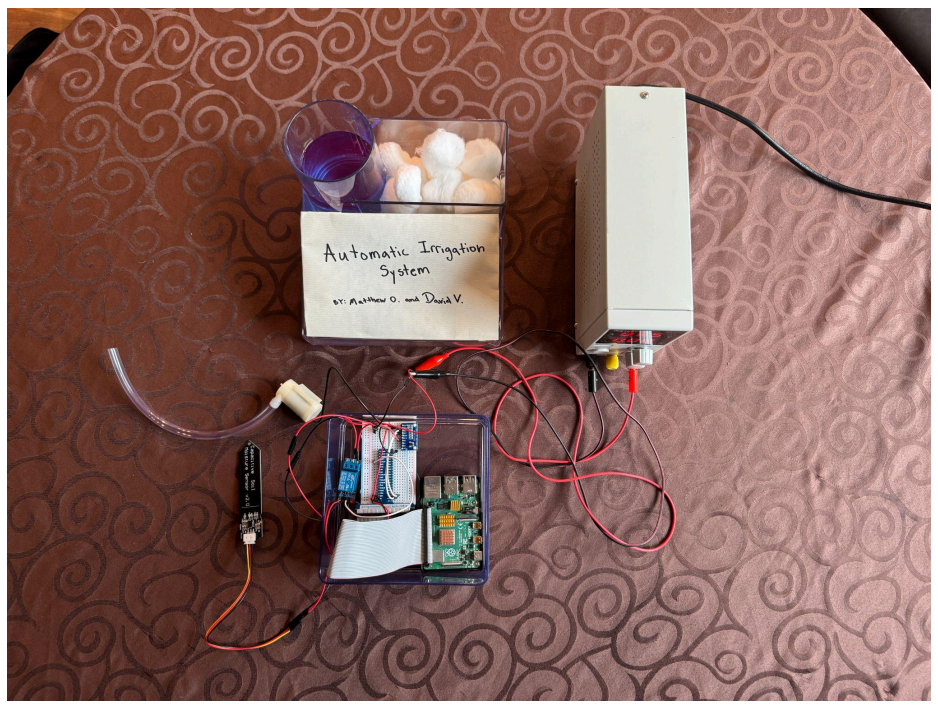

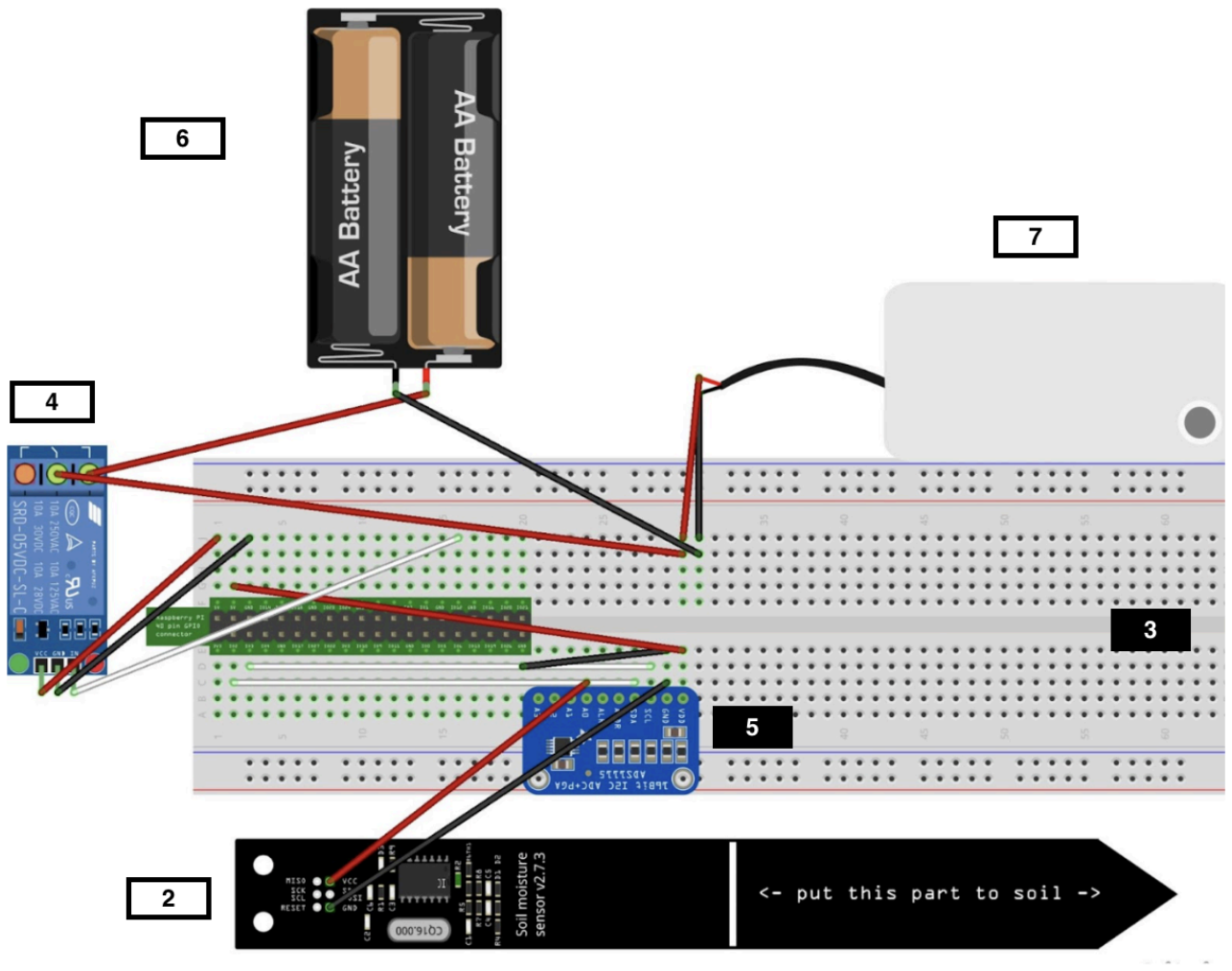

## LINK TO DEMONSTRATION:

**https://youtu.be/xho18kHI4dE**

## VII. PARTS USED

1. Raspberry Pi 4B
2. Capacitive Moisture Sensor
    a. https://gikfun.com/products/gikfun-capacitive-soil-moisture-sensor-corrosion-resistant-for-arduino-moisture-detection-garden-watering-diy-pack-of-2pcs
3. Breadboard
4. 5V Single-Channel Relay Module
    a. https://components101.com/switches/5v-single-channel-relay-module-pinout-features-applications-working-datasheet
5. ADS1115 16-Bit ADC 4 Channel
    a. https://www.addicore.com/products/ads1115-16-bit-adc-4-channel?srsltid=AfmBOorRaN3ntVQyFt3fsATZc_cy_UVdxy1b62i7yLaa1oaAXNg4UO9I
6. 5V Battery
7. Water Pump
    a. https://www.amazon.com/Miokycl-Submersible-Aquarium-JT-DC3L-Vertical/dp/B0C7VTSBMF?th=1

**Specification[4]:**

Capacitive Soil Moisture Sensor:
-Operating voltage: 3.3 ~ 5.5 VDC
-Output voltage: 0 ~ 3.0 VDC
-Interface: PH2.54-3P
-Size: 98 x 23mm (LxW)
-Pin: Analog signal output, GND, VCC

1 Channel 5V Relay Module:
-VCC: Connect 5V Positive Pole Power Supply
-GND: Connect 5V Negative Pole Power Supply

Mini Water Pump:
-DC Voltage: 3-5 V
-Outside diameter of water outlet: 0.29"/7.5mm
-Inside diameter of water outlet: 0.17"/4.5mm
-Water inlet diameter: 0.19"/5mm
-Driving mode: brushless dc design, magnetic driving
-Continuous working life of 300 hours

Vinyl Tubing:
-Material: PVC
-I.D. Size: 0.22"/5.54mm
-O.D. Size: 0.32"/8.20mm

**ADC Specifications**

Resolution & Package: 16-bit analog-to-digital converters in ultra-small QFN
(2 mm × 1.5 mm × 0.4 mm) or MSOP-10 packages.

Supply & Power: Operate from a single 2.0 V to 5.5 V supply; low current draw (150 µA in continuous mode, automatic shutdown in single-shot mode).

Performance: Programmable data rates from 8 SPS to 860 SPS; onboard low-drift voltage reference and internal oscillator.

Integrated Features:

- PGA (ADS1114/ADS1115) allows input ranges down to ±256 mV.
- Input Multiplexer (ADS1115) offers four single-ended or two differential inputs.
- I²C Interface supports four selectable addresses.
- Programmable Comparator (ADS1114/ADS1115).

Operating Modes: Continuous conversion or single-shot (power-down after each reading).

Temperature Range: –40°C to +125°C.

Typical Applications: Battery monitoring, temperature measurement, portable instrumentation, consumer goods, and factory automation.

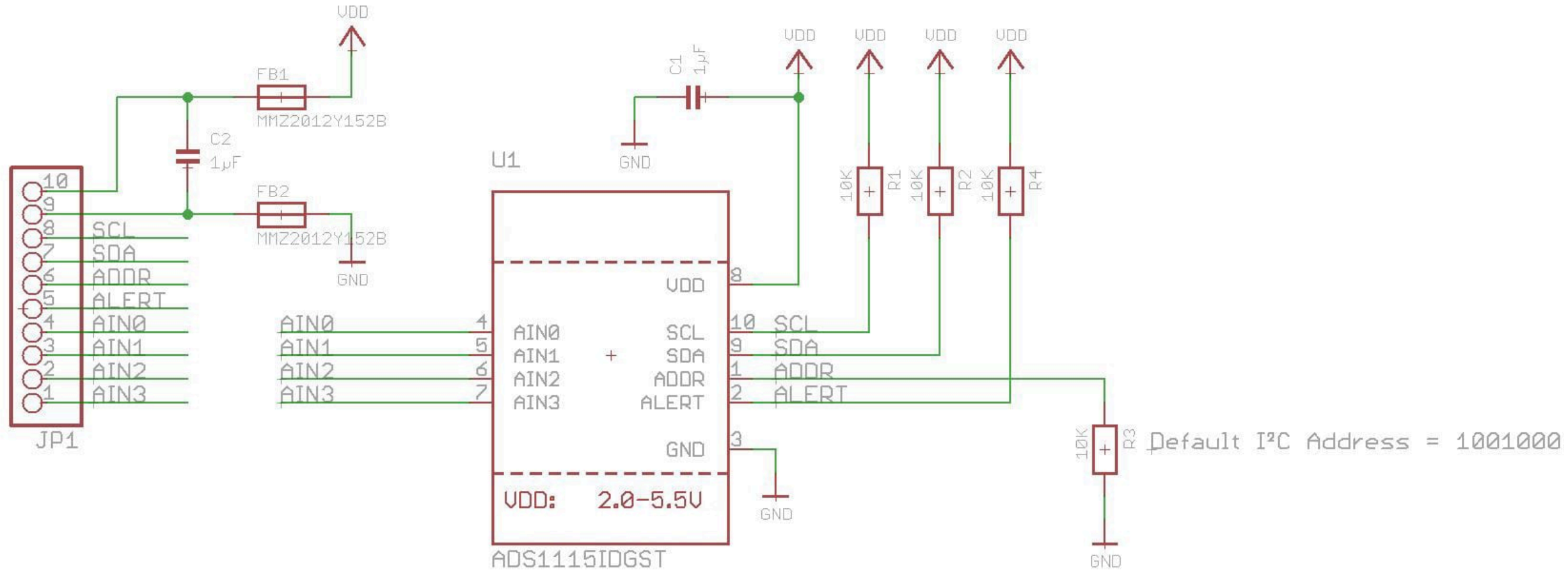

Reference: Texas Instruments[5]

**IX. CONCLUSION** This project demonstrates the potential of an automated irrigation system to conserve water and promote efficient plant care. The combination of reliable hardware, robust algorithms, and an intuitive interface makes it a practical solution for a variety of applications. Future enhancements could include integrating weather forecasts and expanding the system to support multiple plants.